\documentclass[final,3p,times,12pt]{elsarticle}

\usepackage{multirow,setspace,times,amssymb,amsmath,graphicx,color,rotating,subfigure,url}
\usepackage{lineno,color}
\usepackage{natbib}
\usepackage{booktabs}%
\usepackage{longtable}
\graphicspath{{Figures/}}
\usepackage[table]{xcolor}
\usepackage{tabularx}
\usepackage{graphicx} 
\usepackage{epstopdf}
\usepackage{mathrsfs}
\usepackage{makecell}
\usepackage[bookmarks=true,colorlinks,linkcolor=blue,anchorcolor=blue,citecolor=blue,unicode]{hyperref}
\usepackage{bookmark}
\usepackage{setspace}

\hypersetup{CJKbookmarks=true}%
\journal{Eur. Phys. J. B}
\begin{document}

\begin{frontmatter}
\title{Impact of shocks to economies on the efficiency and robustness of the international pesticide trade networks}

\author[SB]{Jian-An Li}
\author[SP,RCE]{Li Wang}
\author[SB,RCE]{Wen-Jie Xie\corref{WJ}}
\ead{wjxie@ecust.edu.cn}
\author[SB,RCE,DM]{Wei-Xing Zhou}
\cortext[WJ]{Corresponding author. Corresponding to: 130 Meilong Road, P.O. Box 114, School of Business, East China University of Science and Technology, Shanghai 200237, China.}

\address[SB]{School of Business, East China University of Science and Technology, Shanghai 200237, China}
\address[SP]{School of Physics, East China University of Science and Technology, Shanghai 200237, China}
\address[RCE]{Research Center for Econophysics, East China University of Science and Technology, Shanghai 200237, China}
\address[DM]{School of Mathematics, East China University of Science and Technology, Shanghai 200237, China}

\begin{abstract}
Pesticides are important agricultural inputs to increase agricultural productivity and improve food security. The availability of pesticides is partially achieved through international trade. However, economies involved in the international trade of pesticides are impacted by internal and external shocks from time to time, which influence the redistribution efficiency of pesticides all over the world. In this work, we adopt simulations to quantify the efficiency and robustness of the international pesticide trade networks under shocks to economies. Shocks are simulated based on nine node metrics, and three strategies are utilized based on descending, random, and ascending node removal. It is found that the efficiency and robustness of the international trade networks of pesticides increased for all the node metrics except the clustering coefficient. Moreover, the international pesticide trade networks are more fragile when import-oriented economies are affected by shocks.
\end{abstract}


\end{frontmatter}

\section{Introduction}
\label{intro} 

Pesticides include insecticides, fungicides, herbicides, disinfectants, and rodenticides and other similar products, which are invented to protect agricultural crops from harms caused by insects, fungi, weeds, viruses, and rats. Therefore, the main functions of pesticides are to reduce yield losses, regulate plant growth, and increase agricultural productivity, which is essential to improving food security. The availability of pesticides in most economies is partially supported by international trade. However, there are internal and external shocks from time to time that affect the involved economies and their ability to trade pesticides internationally. Such shocks to economies influence the efficiency of pesticides redistribution all over the world. It is, thus, important to quantify the efficiency and robustness of the international pesticide trade networks (iPTNs).

The structural properties and their evolutionary behavior of the iPTNs have been studied \cite{Li-Xie-Zhou-2021-FrontPhysics,Li-Wang-Xie-Zhou-2023-Heliyon}. Moreover, the structural robustness and efficiency-based robustness of the iPTNs have also been investigated when the trade relationships are affected by internal and external shocks \cite{Xie-Li-Wei-Wang-Zhou-2022-SciRep}. When the economies involved in the international pesticide trade are influenced by shocks, the structural robustness of the iPTNs has been quantified \cite{Li-Wang-Xie-Zhou-2023-JManagSciEngin}. In this work, we aim to complete the unfinished puzzle by quantifying the efficiency and efficiency-based robustness of the iPTNs under shocks to economies. We note that the efficiency-based robustness of complex networks are less studied \cite{Oehlers-Fabian-2021-Mathematics,Schaeffer-Valdes-Figols-Bachmann-Morales-BustosJimenez-2021-JComplexNetw}. 

Concerning the structural robustness of most complex networks, accumulating evidence shows that complex networks are robust to internal shocks (or random failures) but may be fragile to external shocks (or intentional attacks) \cite{Albert-Jeong-Barabasi-2000-Nature,Cohen-Erez-benAvraham-Havlin-2000-PhysRevLett,Callaway-Newman-Strogatz-Watts-2000-PhysRevLett,Cohen-Erez-benAvraham-Havlin-2001-PhysRevLett}. Researchers have performed extensive analysis of the robustness and fragility of complex networks in various fields, such as international oil trade networks \cite{Foti-Pauls-Rockmore-2013-JEconDynControl,Fair-Bauch-Anand-2017-SciRep,Xie-Wei-Zhou-2021-JStatMech,Wei-Xie-Zhou-2022-JComplexNetw,Chen-Ding-Zhang-Zhang-Nie-2022-Energy,Wei-Xie-Zhou-2022-Energy}, infrastructure networks \cite{Albert-Albert-Nakarado-2004-PhysRevE,Latora-Marchiori-2005-PhysRevE,Arianos-Bompard-Carbone-Xue-2009-Chaos,Pagani-Aiello-2013-PhysicaA,Wang-Guan-Lai-2009-NewJPhys,Zanin-Lillo-2013-EurPhysJ-SpecTop,Diao-Sweetapple-Farmani-Fu-Ward-Butler-2016-WaterRes,Jo-Gao-Liu-Li-Shen-Xu-Gao-2021-IntJModPhysC}, and the networks of Cosa Nostra affiliates \cite{Musciotto-Micciche-2022-EPJDataSci}, to list a few. 
To carry out simulations for quantifying the efficiency and robustness of the iPTNs under shocks to economies, we utilize nine node metrics (clustering coefficient; betweenness; in-degree, PageRank, authority, and in-closeness; out-degree, hub, and out-closeness) and three node removal strategies (descending, random, and ascending).

The remainder of this work is organized as follows. We briefly describe the data in Section~\ref{S1:Data}. We quantify the network efficiency of iPTNs in Section~\ref{S1:iPTN:Effic:Robust:Nodes:NetEffic} and the efficiency-based robustness of iPTNs in Section~\ref{S1:iPTN:Effic:Robust:Nodes:Robust}. We summarize our results in Section~\ref{S1:Summary}.

\section{Data description}
\label{S1:Data}

The data sets for the international pesticide trade were retrieved from the UN Comtrade database (publicly available at https://comtrade.un.org), which covers the period from 2007 to 2018 and contains the import and export economies and trading values of five categories, including insecticides (380891), fungicides (380892), herbicides (380893), disinfectants (380894), and rodenticides and other similar products (380899).

Based on the data sets, for each year, we construct international pesticide trade networks for each category of pesticides, where involved economies are nodes and a directed link is drawn when two economies trade \cite{Xie-Wei-Zhou-2021-JStatMech,Xie-Li-Wei-Wang-Zhou-2022-SciRep}.

\section{Network efficiency}
\label{S1:iPTN:Effic:Robust:Nodes:NetEffic}

\subsection{Shortest path length between economies}

\begin{figure*}[!t]
\centering
    \includegraphics[width=0.32\linewidth]{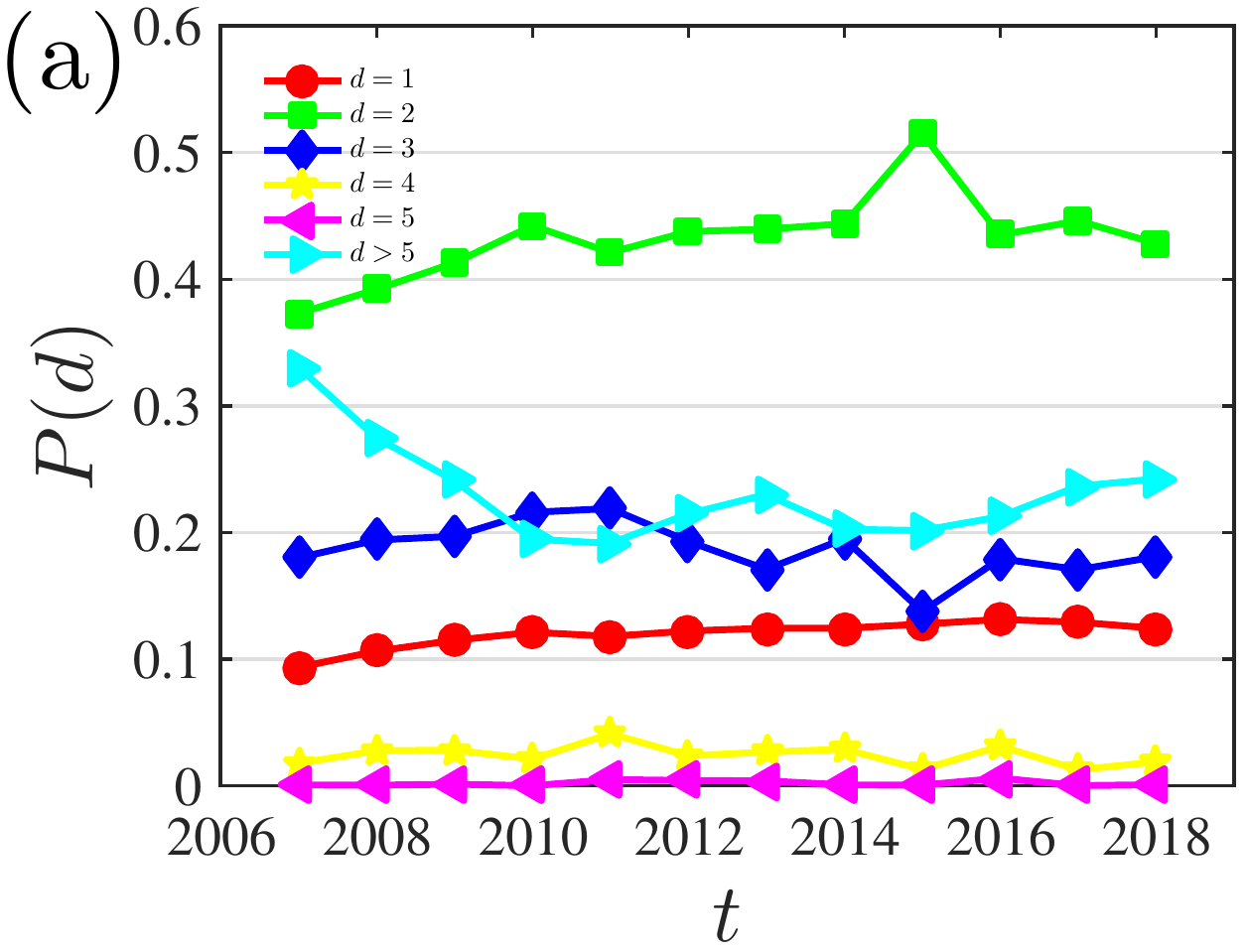}
    \includegraphics[width=0.32\linewidth]{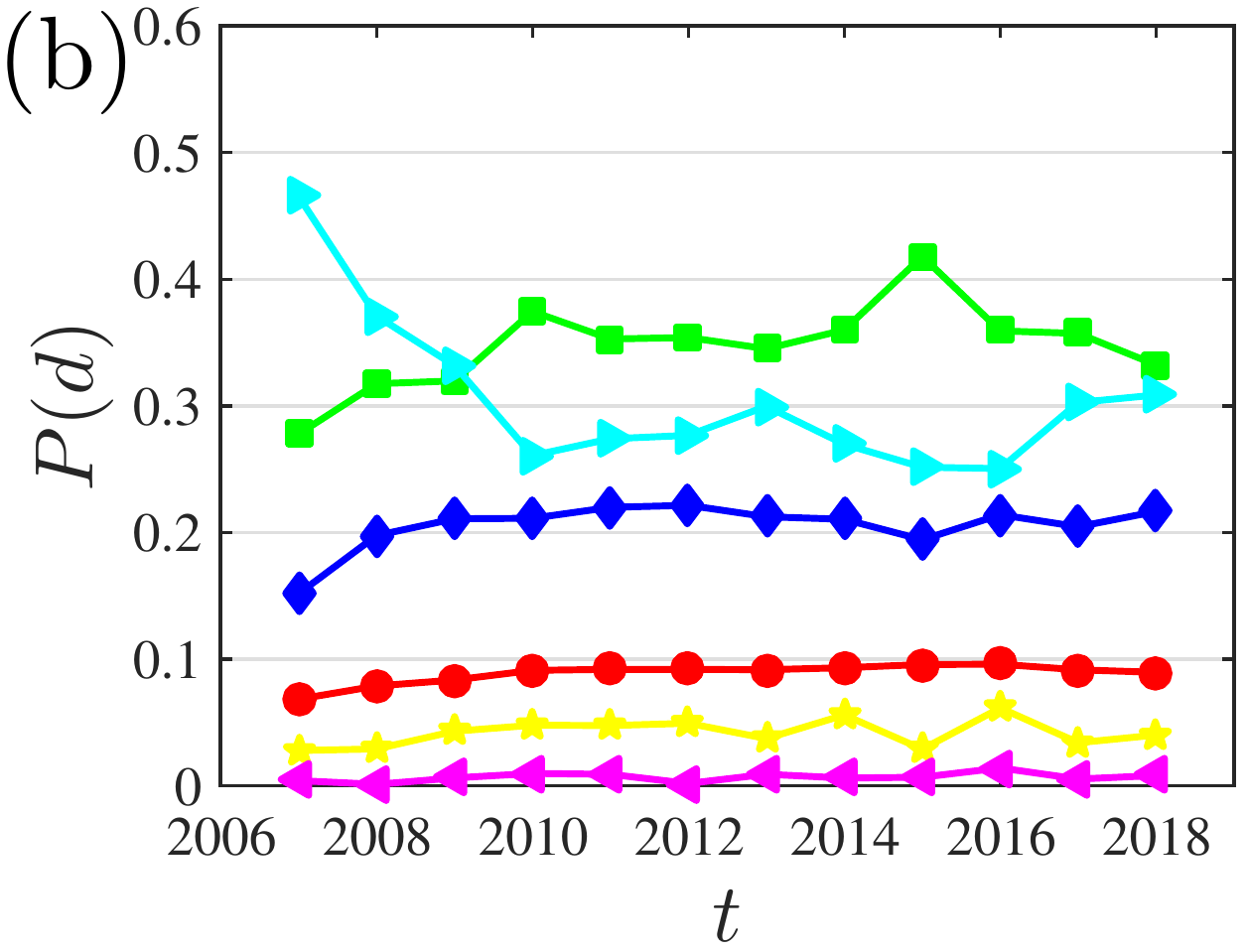}
    \includegraphics[width=0.32\linewidth]{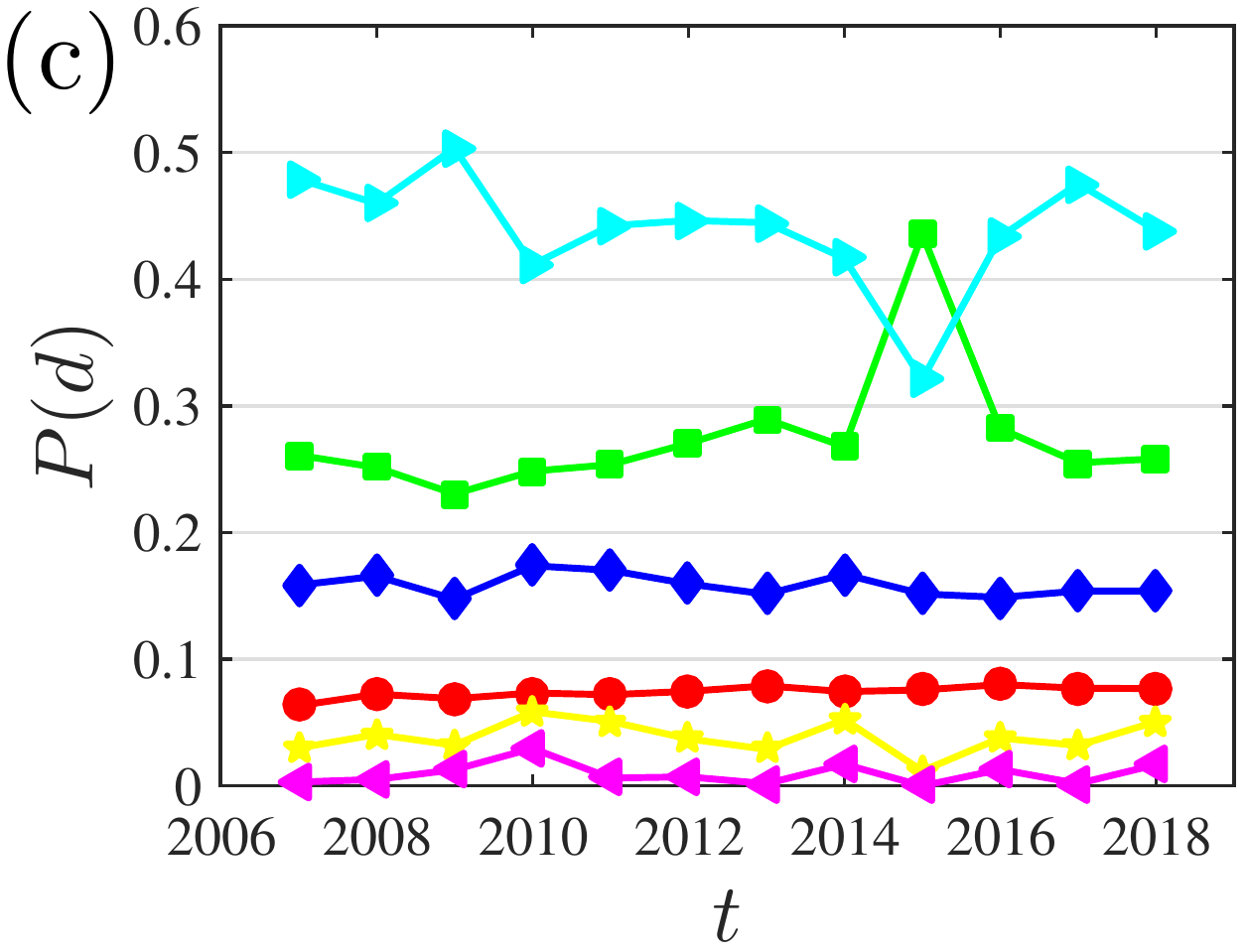}
    \includegraphics[width=0.32\linewidth]{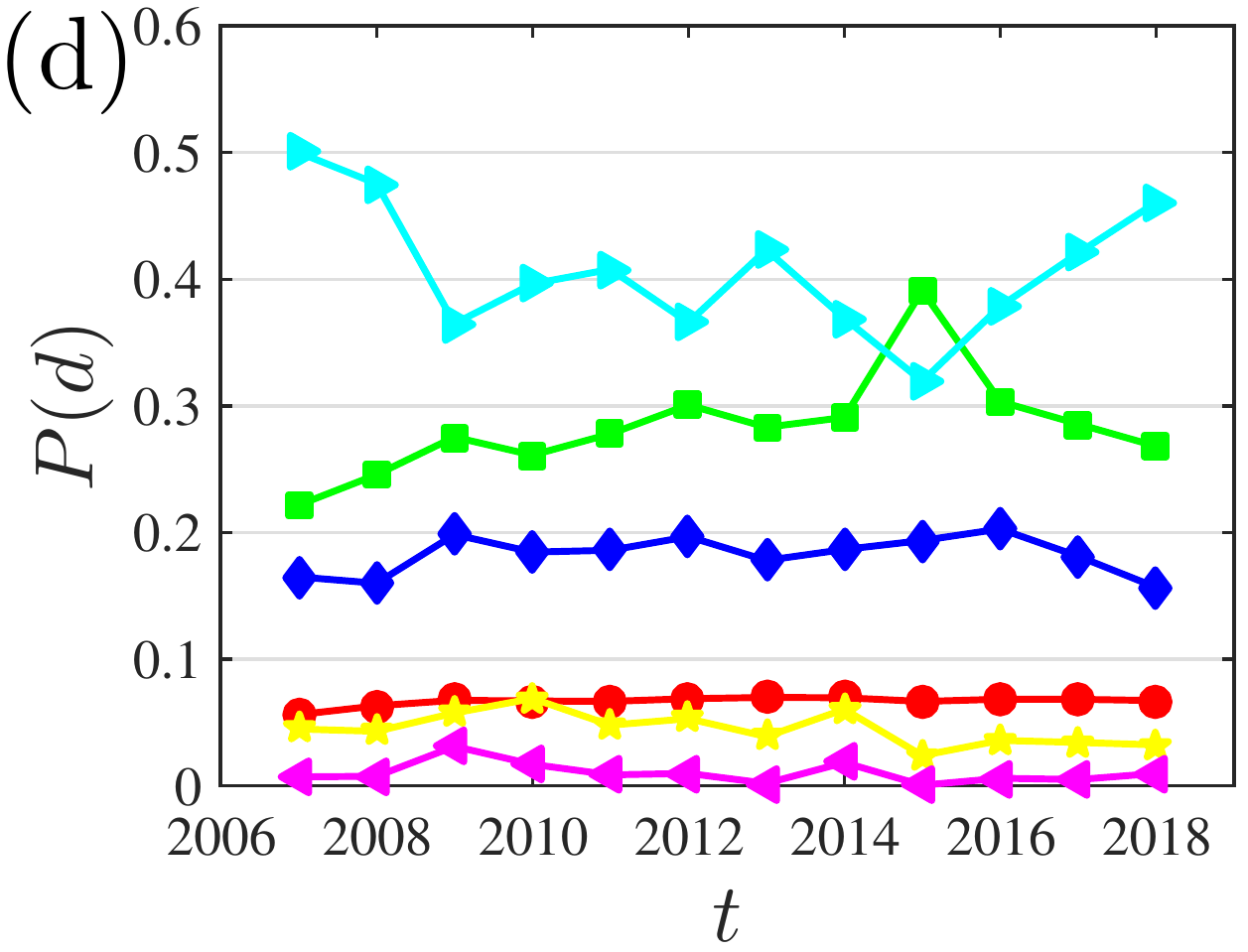}
    \includegraphics[width=0.32\linewidth]{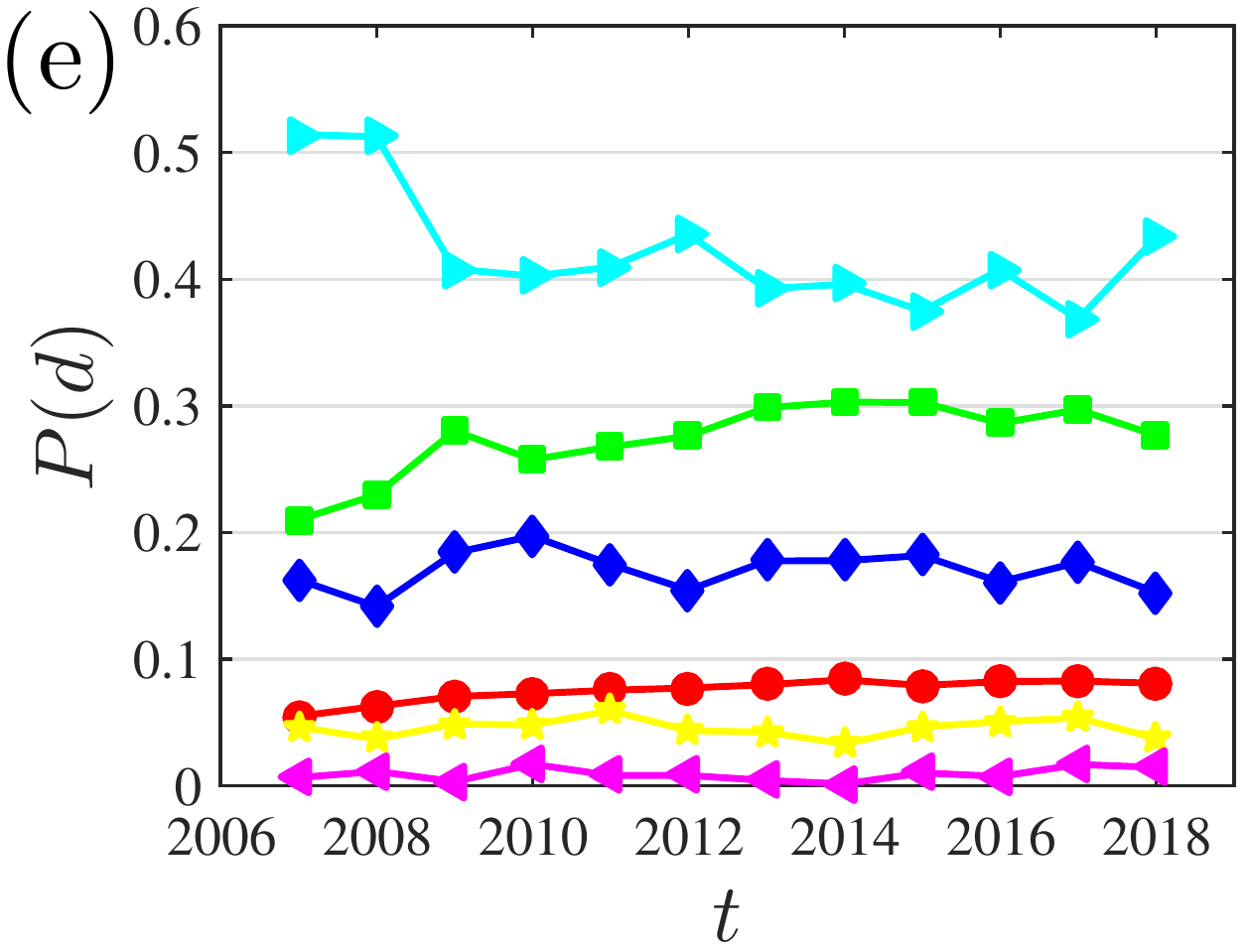}
    \includegraphics[width=0.32\linewidth]{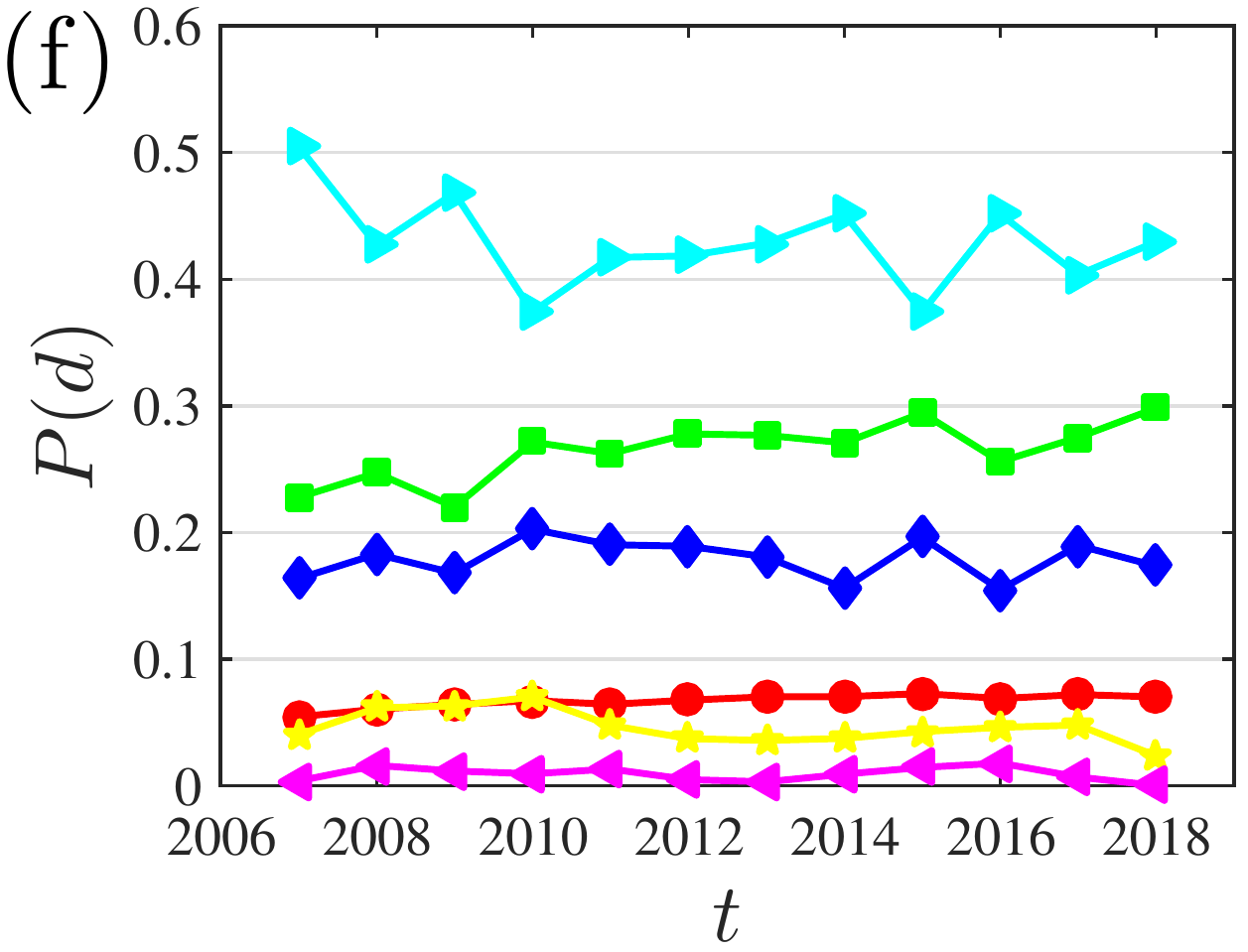}
  \caption{
Distance distribution among economies in the international pesticide trade network. $d$ is the network distance between economies in the directed pesticide trade network. $P(d)$ represents the proportion of the shortest path with a distance of $d$. The figure shows the time evolution of the proportion of the distance $d=1$, 2, 3, 4, 5, and $d>5$. The six plots correspond to the trade networks of pesticide products, including (a) aggregated, (b) insecticides, (c) fungicides, (d) herbicides, (e)  disinfectants, and (f) rodenticides and other similar products.}
    \label{Fig:Pesticide:node:efficiency:SP}
\end{figure*}

In the international pesticide trade networks, pesticide products flow between economies. There are a lot of pesticide-producing and pesticide-consuming economies in the trade network, and there are also some economies with a lot of pesticide imports and exports. The network efficiency analysis in this paper is based on directed networks, and the relevant metric indicators are for asymmetric matrices, except for the clustering coefficient. Network efficiency is also based on directed shortest paths. In the international pesticide trade system, different economies play different roles and contribute their own strengths to the global pesticide trade and food production. It is difficult to distinguish between exporting and importing economies because the international pesticide trade networks include a large number of economies that both export and import. Network efficiency refers to the flow efficiency of information, energy, and matter in the network, such as the information transmission efficiency of the information network and the traffic efficiency of the traffic network \cite{Latora-Marchiori-2001-PhysRevLett,Holme-Kim-Yoon-Han-2002-PhysRevE}. Generally speaking, the trade network efficiency can be used as the path length between the producer and the consumer. The larger the path length, the lower the efficiency; otherwise, the higher the efficiency.

We calculated the shortest path length between economies in the iPTNs. Figure~\ref{Fig:Pesticide:node:efficiency:SP} shows the evolution of occurrence frequencies $P(d)$ for $d=1$, $d=2$, $d=3$, $d=4$, $d=5$, and $d>5$, where $d$ is the distance between the economies in the pesticide trade network.

Overall, the occurrences of shortest path length in the six plots are relatively similar, especially in plot (c) for fungicides, plot (d) for herbicides, plot (e) for disinfectants, and plot (f) for rodenticides and other similar products. In each of these four temporal iPTNs, the proportion $P(d)$ with $d > 5$ is basically the largest, which is close to 0.5 in 2007, but most of them have an infinite shortest path length. It is due to the fact that the iPTNs we are studying are directed and thus there is no reachable path between some economies. With the development of globalization, there are more and more trade relations, and the proportion of economy pairs with infinite shortest path lengths is getting smaller and smaller until around 2015. The main reason is that the sales in the international pesticide market plunged 8.5\% year on year, the steepest decline in more than a decade. 
In most years, we have
\begin{equation}
  P(2)>P(3)>P(1)>P(4)>P(5).
\end{equation}
This relation does not hold for the aggregated iPTN in Fig.~\ref{Fig:Pesticide:node:efficiency:SP}(a) and the iPTN for insecticides in Fig~\ref{Fig:Pesticide:node:efficiency:SP}(b). The aggregated iPTN integrates the five pesticide products and has the largest number of links. Moreover, compared with the other four pesticide trade products, the insecticides trade network showed the largest trade volume and the largest trade relationships.

The proportion $P(1)$ of distance $d=1$ in Fig.~\ref{Fig:Pesticide:node:efficiency:SP} measures the density of the pesticide trade network. It can be seen from the figure that, with the evolution of time $t$, $P(1)$ becomes larger and larger, indicating that there are more and more pesticide trade relations, but after 2015, there is also a downward trend.

The proportion $P(2)$ of distance $d=2$ measures that the economies in the pesticide trade network do not have direct trade relations, but have indirect trade relations through intermediate economies. It can be seen from Fig.~\ref{Fig:Pesticide:node:efficiency:SP} that, with the evolution of time $t$, $P (2)$ becomes larger and larger, indicating that there are more and more pesticide trade relations, but after 2015, there is also a downward trend. In particular, the $P(2)$ curves for the aggregated iPTN in Fig.~\ref{Fig:Pesticide:node:efficiency:SP}(a), the iPTN of insecticides in Fig.~\ref{Fig:Pesticide:node:efficiency:SP}(b), and the iPTN of fungicides in Fig.~\ref{Fig:Pesticide:node:efficiency:SP}(c) showed obvious fluctuations in 2015.

Generally speaking, the shorter the trade distance between economies in the international trade network, the higher the trade efficiency. However, the international trade network is affected by geographical location, political factors, resource endowment distribution, etc. It is impossible that the shortest path length of the trade network is too small. Therefore, we introduced the index of network efficiency to measure the trade flow efficiency and the change rule of the trade structure of pesticide products in the international pesticide trade network.

\subsection{Network efficiency}

Network efficiency is defined as the average reciprocal path length between nodes in the network. The path length between nodes is defined differently between directed networks and undirected networks, as well as between weighted networks and unweighted networks. In an unweighted, directed network, the network efficiency $E$ is defined as follows \cite{Latora-Marchiori-2001-PhysRevLett,Holme-Kim-Yoon-Han-2002-PhysRevE},
\begin{equation}\label{Eq:weight:efficiency}
  E=\frac1{N(N-1)}\sum_{i\neq j}e_{ij},
\end{equation}
where $e_{ij}$ represents the path efficiency between economy $i$ and economy $j$, measuring the transmission efficiency of material flow, information flow, or energy flow between two nodes, specifically defined as
\begin{equation}
  e_{ij}=\frac1{d_{ij}},
\end{equation}
where $d_{ij}$ represents the shortest path length between economy $i$ and economy $j$.

\begin{figure}[!h]
\centering
    \includegraphics[width=0.49\linewidth]{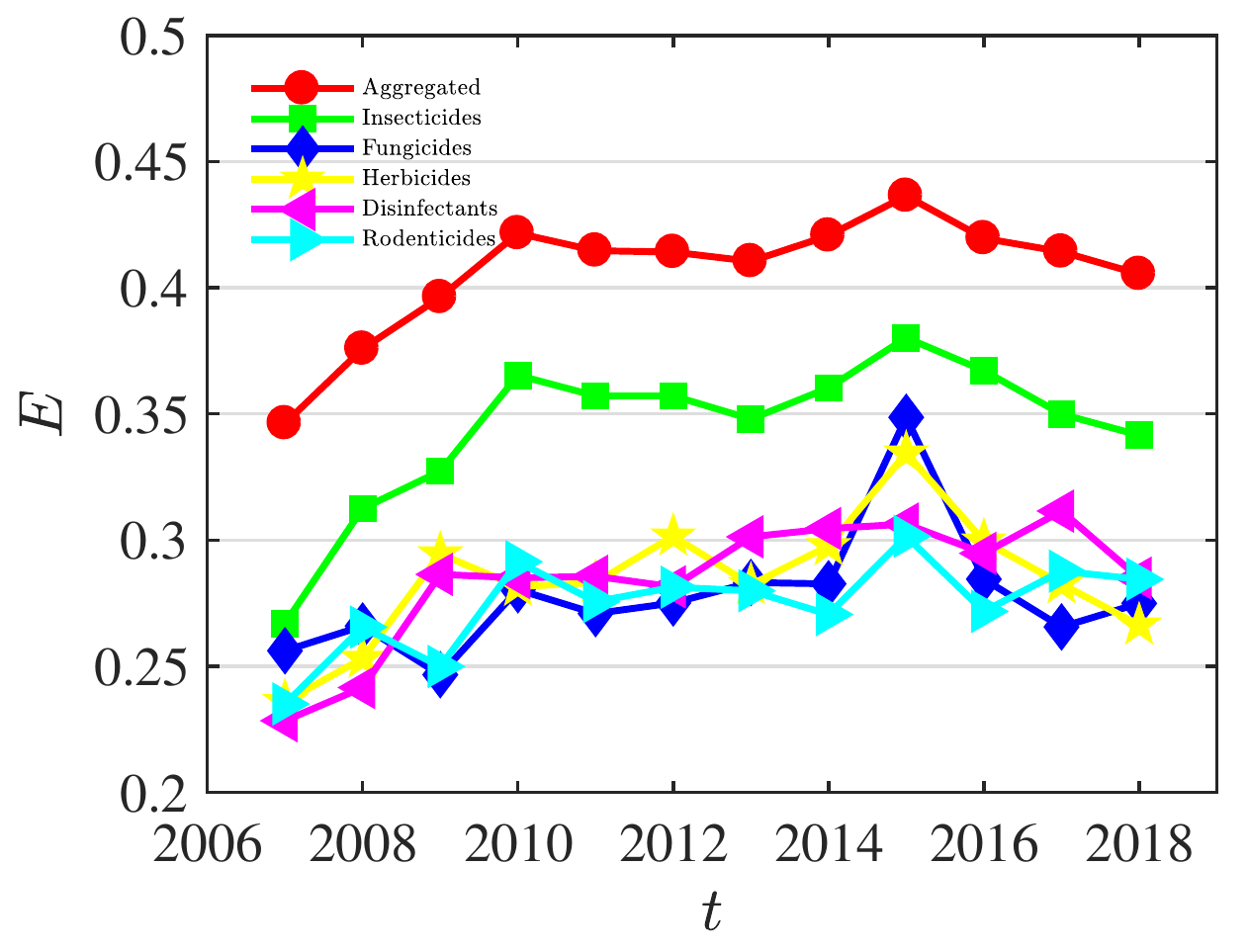}
  \caption{Evolution of the efficiency of the international pesticide trade networks. Each solid line corresponds to an international pesticide trade network, including aggregated, insecticides, fungicides, herbicides, disinfectants and rodenticides and other similar products.}
    \label{Fig:Pesticide:node:efficiency:Def}
\end{figure}

Fig.~\ref{Fig:Pesticide:node:efficiency:Def} shows the evolution of the efficiency of the international pesticide trade networks. Each solid line corresponds to an international pesticide trade network, including aggregated, insecticides, fungicides, herbicides, disinfectants and rodenticides and other similar products. The five types of pesticide trade network efficiency have similar evolutionary patterns, which can be basically divided into three stages. The first stage is from 2007 to 2010, which can be called the rapid growth period of network efficiency. The period from 2010 to 2015 can be regarded as a stable period, and the efficiency of the international pesticide trade network has not changed significantly. After 2015, it can be seen as a period of declining network efficiency. The efficiency of the international pesticide trade network has declined.

\begin{table*}[!th]
\centering
\caption{The nine node metrics used in this work.}
\begin{tabular}{p{2.35cm} p{1.2cm} p{10cm} p{1cm}}
\hline
Variable & Definition & Description & References \\
\hline
In-degree   & $ \sum_{i=1}^{N}a_{ij}$ & $N$ is the number of nodes in the trade network. $a_{ij}$ is the element of adjacent matrix of trade network. & \cite{Wei-Xie-Zhou-2022-Energy}   \\
Out-degree  & $ \sum_{j=1}^{N}a_{ij}$ & It is the count of importing partners of economy $i$. & \cite{Wei-Xie-Zhou-2022-Energy}   \\
In-closeness  & $\frac{N-1}{\sum_{j\in{{\mathbf{n}}_i^{\mathrm{in}}}}l_{ij}}$ & It is defined as the reciprocal of the mean distance of $i$ to all other nodes. & \cite{Wei-Xie-Zhou-2022-Energy}   \\
Out-closeness   & $\frac{N-1}{\sum_{j\in{{\mathbf{n}}_i^{\mathrm{out}}}}l_{ij}}$ & It is defined as the reciprocal of the mean distance of all other nodes to $i$. & \cite{Wei-Xie-Zhou-2022-Energy}   \\
Authorities/Hubs  & - & The authority score and hub score of a node can be calculated in an iterative way. & \cite{Kleinberg-1999-JACM}   \\
PageRank  & - & It is an important variant of eigenvector centrality. & \cite{Brin-Page-1998-ComputNetwISDNSyst}   \\
Betweenness  & $ \sum_{st}\frac{n_{st}^i}{g_{st}}$ & It is the ratio of the number of shortest paths go through the investigated node to the number of shortest paths between all pairs of nodes in the network. & \cite{Freeman-1977-Sociometry}   \\
Clustering   & $ \frac{2n_i}{k_i(k_i-1)}$ & It characterizes the connectance of its trade partners. $n_i$ is the number of triangles adjacent to $i$.  & \cite{Watts-Strogatz-1998-Nature}   \\
\hline
\end{tabular}
\label{Table:node:metrics}
\end{table*}

\section{Network robustness against shocks to economies}
\label{S1:iPTN:Effic:Robust:Nodes:Robust}

\subsection{Node metrics}

There are many node metrics capturing the local or global characteristics of nodes that can be used to measure the importance of nodes in different aspects \cite{Chen-Lu-Shang-Zhang-Zhou-2012-PhysicaA,Lu-Chen-Ren-Zhang-Zhang-Zhou-2016-PhysRep}. We consider here nine node metrics, including clustering coefficient, betweenness, PageRank, in-degree, out-degree, in-closeness, out-closeness, authorities, and hubs, as shown in Table~\ref{Table:node:metrics}. These node metrics have different traits. The clustering coefficient is defined for undirected networks, while the other eight metrics are extracted from directed networks. Concerning the other eight node metrics, betweenness does not distinguish between importing and exporting economies; in-degree, PageRank, authority, and in-closeness are calculated from exporting (thus the node under consideration is a target node); and out-degree, out-closeness, and hub are obtained from importing economies (thus the node under consideration is a source node). In this section, we quantify the importance of the economy based on these nine node metrics and further investigate their mutual correlations.

\subsection{Network efficiency under shocks to economies}

Complex network structures and network functions are closely related, and different network structures have different roles and functions for specific problems and backgrounds. In the analysis of network stability, indicators such as in-degree, authorities, and betweenness have a great impact on the network structure, but not all indicators are equally important for all network functions. Therefore, we analyzed the differences and impacts of important indicators in the study of network efficiency. Network efficiency describes the transmission efficiency of material flow, information flow, or energy flow in a complex network, which has important research and application value.

To compare the changes in network efficiency after removing network nodes according to the indicators of different nodes, we take the network efficiency of the original network as the benchmark, recalculate the efficiency of the residual network after removing network nodes, and calculate the ratio between the two so that we can horizontally compare the changes in network efficiency of different pesticide trade networks. The formula of the ratio $\beta^I$ is
\begin{equation}
   \label{Eq:Efficiency:Measure}
   \beta^I(p)=\frac{E_{p}}{E},
\end{equation}
where $E$ represents the network efficiency of the original network, and $E_p$ represents the network efficiency after removing the nodes with a proportion of $p$ according to the given indicators.

We consider three node removal rules. Based on the node metric index $I$, we remove the economies from the pesticide trade network and recalculate the network efficiency for the remaining networks. We then compare the results of the three node removal strategies. The ``descend'' removal rule is to remove a proportion of $p$ of the largest trading economies based on indicator $I$ and recalculate the network efficiency $E_p$ of the remaining network. The ``ascend'' removal rule is to remove a proportion of $p$ of the economies with the smallest $I$. The ``random'' removal method is to randomly remove a proportion of $p$ of the trading economies and then calculate the network efficiency $E_p$ of the remaining network.
In the process of removing network nodes, in addition to randomly removing nodes, node removal is mostly performed in descending order of indicators. We introduce the removal of indicators in ascending order, the main purpose of which is to compare with the other two removal methods. Not all indicators meet the rule that the higher the value, the greater the impact on network efficiency. Considering both ascending and descending removal methods, it is not easy to miss some key indicators. For example, the presence of weak ties during edge removal plays a key role in network connectivity.

\begin{figure*}[!ht]
\centering
    \includegraphics[width=0.89\linewidth]{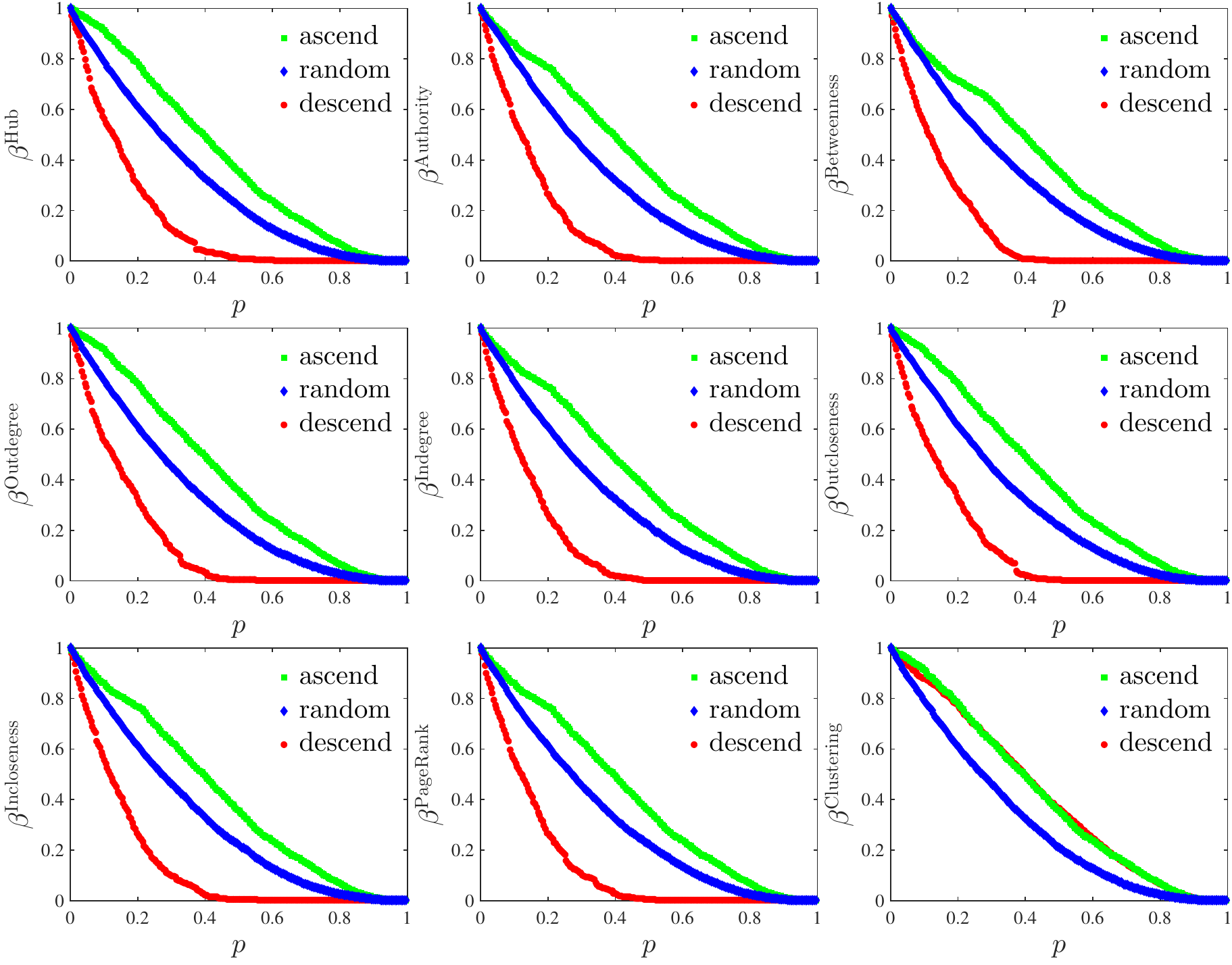}
  \caption{Analysis of network efficiency under shocks to the economies for the aggregated iPTN in 2018. The lines are the $\beta^I$ curves calculated from the largest connected subnetworks after removing nodes based on indicator $I$. The words ``descend'', ``ascend'', and ``random'' in the legend correspond to the cases where the nodes with largest $I$, smallest $I$, and random $I$ are removed preferentially. For the random node removal stategy, we repeated 20 runs and presented the averages.}
    \label{Fig:Pesticide:node:fficiency:GC}
\end{figure*}

For the aggregated iPTN in 2018, we illustrate in Fig.~\ref{Fig:Pesticide:node:fficiency:GC} the $\beta^I(p)$ curves for the nine indicators $I$. The curves associated with ``descend'', ``ascend'', and ``random'' in the figure correspond to the priorities of removing nodes with the largest $I$ values, the smallest $I$ values, and random $I$ values, respectively. We find that the results are similar for different node metrics, except for the clustering coefficient. For the clustering coefficient, the two $\beta^I$ curves corresponding to the descending and ascending node removal strategies almost overlap, while the $\beta^I$ curve corresponding to the random node removal strategy decreases faster than the other two curves. For the other eight node metrics, the corresponding $\beta^I$ curves have similar patterns. Specifically, the curves obtained from the descending node removal strategy decrease the fastest, while those from the ascending node removal strategy decrease the lowest. For the descending node removal strategy, the $\beta^I(p)$ value is almost nil when the economies are affected by around 40\%. In comparison, the giant component still contains more than 40\% nodes \cite{Li-Wang-Xie-Zhou-2023-JManagSciEngin}.

\begin{figure*}[!ht]
\centering
    \includegraphics[width=0.89\linewidth]{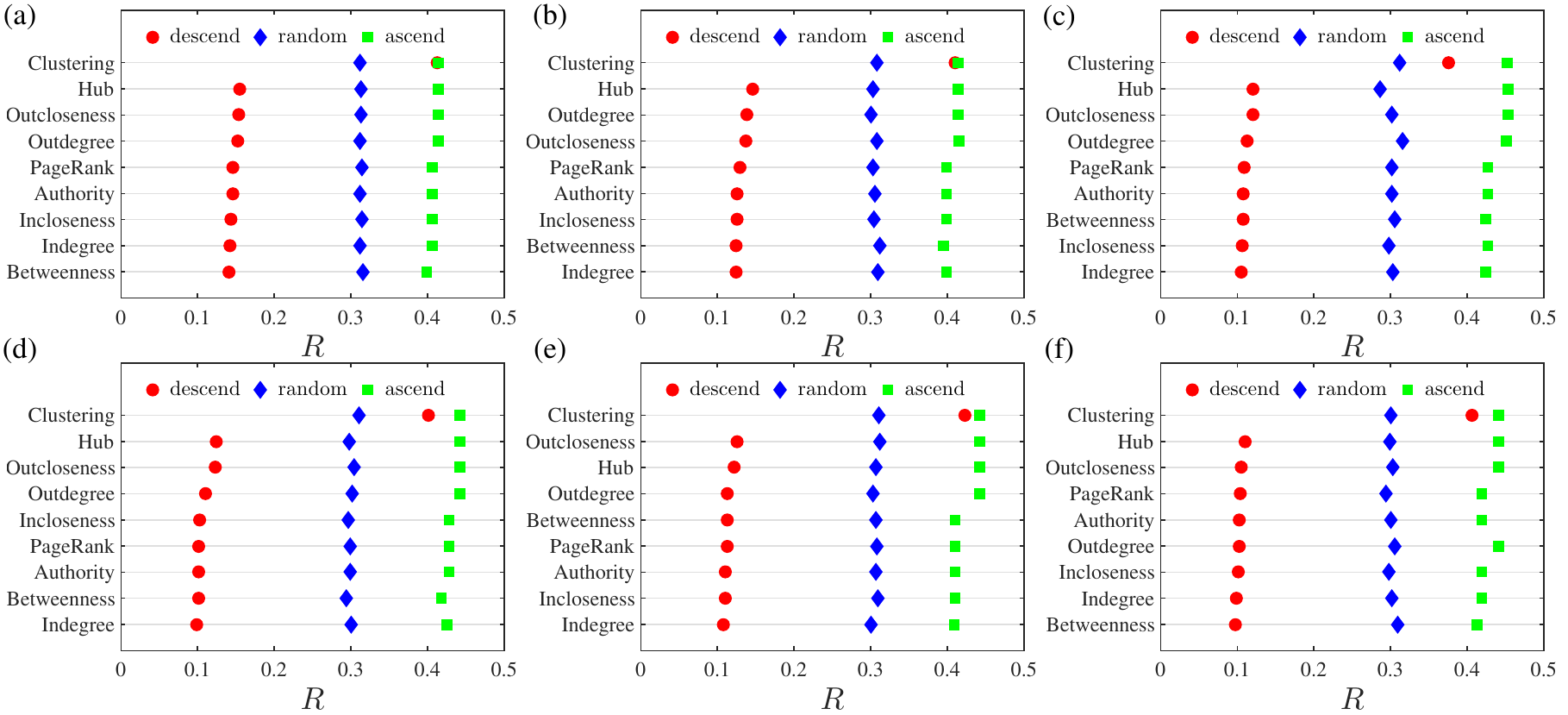}
  \caption{Network efficiency based robustness of the aggregated iPTN (a), the iPTN of insecticides (b), the iPTN of fungicides (c), the iPTN of herbicides (d), the iPTN of disinfectants (e), and the iPTN of rodenticides and other similar products (f) in 2018. The words ``descend'', ``ascend'', and ``random'' in the legend correspond to the cases where the nodes with largest $I$, smallest $I$, and random $I$ are removed preferentially. For the random node removal strategy, we repeated 20 runs and presented the averages.}
    \label{Fig:Pesticide:node:efficiency:R}
\end{figure*}

\subsection{Network-efficiency-based robustness}

To compare the differences in network efficiency under different node removal strategies in a more detailed way, a more quantitative method is also used to analyze the network efficiency. There are many similar studies on the robustness and fragility of complex networks, such as Ref.~ \cite{Cohen-Erez-benAvraham-Havlin-2000-PhysRevLett}. The approach in Ref.~ \cite{Cohen-Erez-benAvraham-Havlin-2000-PhysRevLett} is based on percolation theory and analyzes the general condition for the critical fraction of nodes. Our approach is based on efficiency and robustness as
presented in Ref.~\cite{Schneider-Moreira-Andrade-Havlin-Herrmann-2011-ProcNatlAcadSciUSA}. The area under a curve in Fig.~\ref{Fig:Pesticide:node:fficiency:GC} represents the impact of the corresponding indicator on the network efficiency. The more the area is, the more obvious the network efficiency will be. This efficiency-based robustness can be expressed as follows \cite{Schneider-Moreira-Andrade-Havlin-Herrmann-2011-ProcNatlAcadSciUSA},
\begin{equation}
   R_{\beta}^{I}=\frac{1}{n}\sum_{k=0}^{n}\beta_{\rm{node}}^{I}\left(p_k\right),
\end{equation}
where $p_k=k/n$ indicates the proportion of economies in the pesticide trade network to be deleted. 

Figure~\ref{Fig:Pesticide:node:efficiency:R} shows the extent to which the efficiency of the international pesticide trade network is affected under the impact of shocks to the economies, where Fig.~\ref{Fig:Pesticide:node:efficiency:R}(a) shows the combined trade network of all five global trade networks of pesticide products, and Fig.~\ref{Fig:Pesticide:node:efficiency:R} (b-f) correspond to the trade networks of five pesticide products. Different markers correspond to different node removal strategies.

As can be seen from Fig.~\ref{Fig:Pesticide:node:efficiency:R}, the strategies that remove economies from the iPTNs in descending order according to the clustering coefficient show different behaviors. For the six networks, we have
\begin{equation}
    R_{\beta,\mathrm{ascend}}^{\mathrm{Clustering}} > R_{\beta, \mathrm{descend}}^{\mathrm{Clustering}} > R_{\beta, \mathrm{random}}^{\mathrm{Clustering}},
\end{equation}
showing that the random node removal strategy has the largest impact on network efficiency. The distinct behavior is mainly reflected in the descending node removal strategy.
In contrast, for the other eight node metrics, we have
\begin{equation}
    R_{\beta,\mathrm{ascend}}^{I} > R_{\beta, \mathrm{random}}^{I} > R_{\beta, \mathrm{descend}}^{I},
\end{equation}
showing that the descending node removal strategy has the largest impact on network efficiency and the ascending node removal strategy has the least impact on network efficiency.

We also observe that the strategies based node metrics associated with import (authority, in-closeness, and in-degree) have a larger impact on network efficiency than those with export (hub, out-closeness, and out-degree). The importing economies have a greater impact on network efficiency. This conclusion is consistent with many studies, such as the study of oil network efficiency and robustness, in which the impact of removing importing economies is greater than that of exporting economies \cite{Xie-Wei-Zhou-2021-JStatMech,Wei-Xie-Zhou-2022-Energy}.

\begin{figure*}[!th]
\centering
    \includegraphics[width=0.89\linewidth]{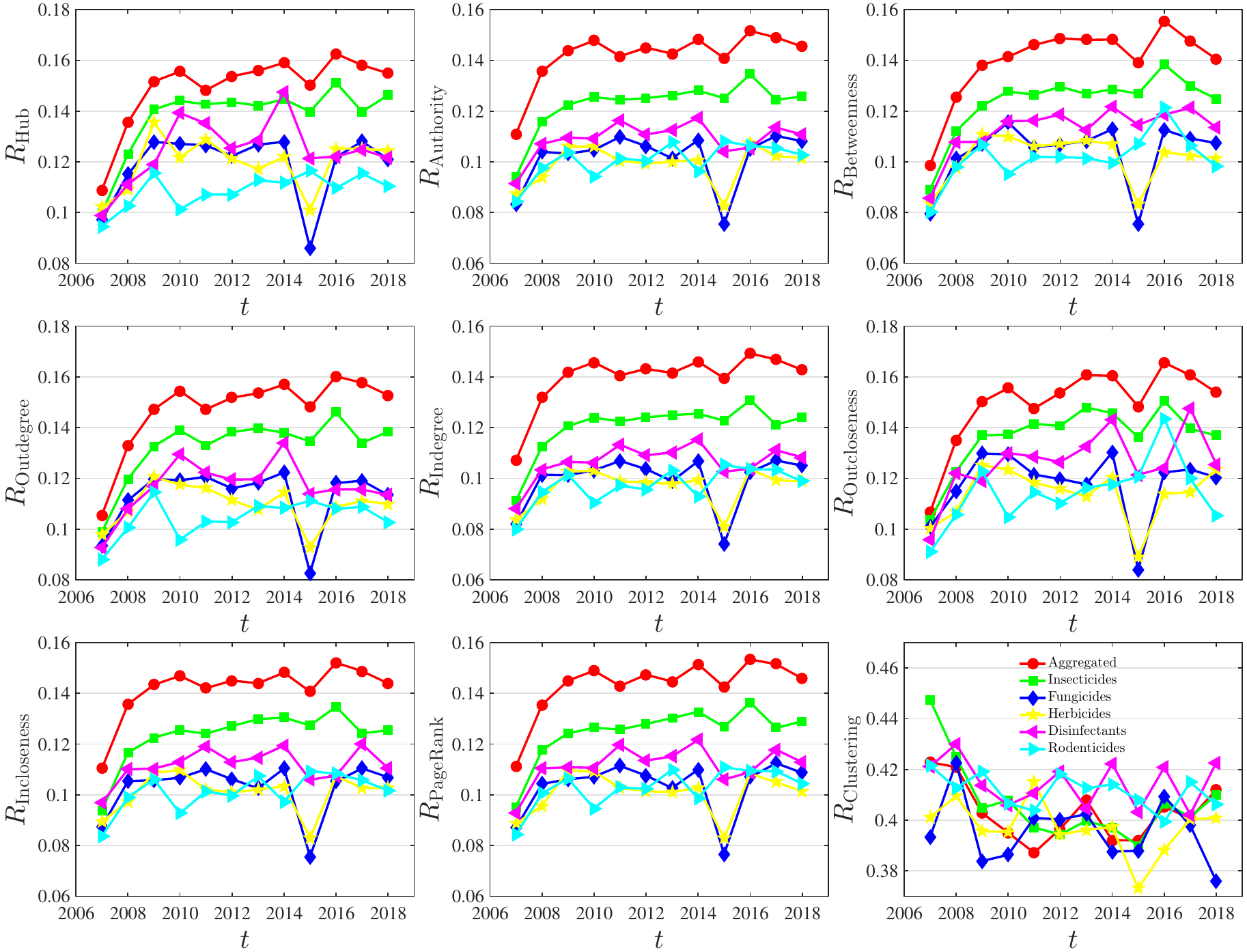}
  \caption{Evolution of the network-efficiency-based robustness of the aggregated iPTN and the iPTNs of insecticides, fungicides, herbicides, disinfectants, and rodenticides and other similar products for the descending node removal strategies based on the nine node metrics.
  }
    \label{Fig:Pesticide:Revo:node:efficiency}
\end{figure*}

\subsection{Evolution of network-efficiency-based robustness}

We now turn to investigate the evolution of the network-efficiency-based robustness of the aggregated iPTN and the five iPTNs of insecticides, fungicides, herbicides, disinfectants, and rodenticides and other similar products from 2007 to 2018 for the descending node removal strategies based on the nine node metrics. The results are presented in Fig.~\ref{Fig:Pesticide:Revo:node:efficiency}. Each plot corresponds to a node metric. Based on this metric, we remove the nodes and then analyze the changes in network efficiency. By comparing the robustness of different networks, we can find that the aggregated network is the most robust. From the definition of network efficiency, it is known that the higher the density of connected edges of the network, the smaller the average path length between nodes, and the higher the network efficiency. Removing the edges of the network has less impact on the connectivity and the average shortest path for a network with a high density of edges, so the corresponding network robustness is higher.

Once more, we can see that while the results for the other eight node metrics are qualitatively more similar, the results for the clustering coefficient are noticeably different. The six robustness curves exhibit an overall decreasing trend. More precisely, the robustness curves decreased in the early years and remained relatively stable with local fluctuations.

In contrast, for the other eight node metrics, the robustness curves show an overall upward trend, which increased in the early years and remained stable. Among the six networks, the aggregated iPTN is the most robust. Among the other five networks, the iPTN of insecticides is the most robust. We also observed a sharp decrease in robustness in 2015, especially for the two iPTNs of herbicides and fungicides. We contend that international pesticide trade networks are becoming more robust in terms of network efficiency in the face of shocks to economies.

\section{Summary}
\label{S1:Summary}

We have investigated the efficiency of the international pesticide networks of insecticides (380891), fungicides (380892), herbicides (380893), disinfectants (380894), and rodenticides and other similar products (380899) from 2007 to 2018, as well as the corresponding aggregated networks of all the five categories of pesticide. We found that the network efficiency increased in the first four years and decreased in the last four years. In addition, for each year, the aggregated iPTN had the highest efficiency and the iPTN of insecticides had the second highest efficiency. These observations are by and large consistent with the time-varying pattern of the number of links \cite{Li-Xie-Zhou-2021-FrontPhysics}. There are, of course, other factors to explore.

We further investigated the robustness of iPTN efficiency, or the efficiency-based robustness of the iPTNs, as adopted in Refs.~\cite{Xie-Wei-Zhou-2021-JStatMech,Xie-Li-Wei-Wang-Zhou-2022-SciRep}. To simulate different types of shocks to economies, we utilized three strategies by removing nodes with descending, random, and ascending orders of nine node metrics. We found that the efficiency-based robustness of the international pesticide trade networks increased for all the node metrics except the clustering coefficient. Moreover, the international pesticide trade networks are more vulnerable when shocks hit import-oriented economies than export-oriented economies. We also found that the aggregated iPTN is the most robust against shocks, and the iPTN of insecticides is the second most robust. The robustness curves are mostly $\cup$-shaped, decorated by a sharp drop in 2015 in some robustness curves, which was caused by an 8 percent plunge in market sales in 2015.

\section*{Acknowledgements}

   This work was supported by the National Natural Science Foundation of China (72171083), the Shanghai Outstanding Academic Leaders Plan, and the Fundamental Research Funds for the Central Universities.
   

\section*{Author contributions}

Funding acquisition: W-XZ; 
investigation: J-AL, LW, W-JX and W-XZ; 
methodology: W-JX and W-XZ; 
supervision: W-JX and W-XZ; 
writing—original draft, J-AL and W-JX; 
writing—review and editing: W-XZ.

\medskip

\noindent
{\small{\bf{Data availability statement}} This manuscript has no associated data or the data will not be deposited. [Authors' comment: The associated data in this manuscript can be retrieved from the UN Comtrade database at {\url{https://comtrade.un.org}}.]}



\end{document}